\documentclass[11pt]{article}
\usepackage[english]{babel}
\usepackage{amsmath,amssymb,amsthm}
\usepackage[all]{xy}

\usepackage[pdfpagemode=UseNone,bookmarks=false,pdfpagelayout=SinglePage,pdfstartview={XYZ null null 1},
pdfhighlight=/P,colorlinks=true,linkcolor=blue,citecolor=blue,urlcolor=blue]{hyperref}

\usepackage[width=0.95\textwidth,indention=0mm,justification=centerlast,
 font=small,labelfont=bf,labelsep=period]{caption}

\allowdisplaybreaks[3]

\advance\textwidth 30mm  \advance\oddsidemargin -15mm
\advance\textheight 36mm \advance\topmargin -24mm

\theoremstyle{plain}
\newtheorem{proposition}{Proposition}
\theoremstyle{remark}
\newtheorem{remark}{Remark}
\DeclareMathOperator{\ad}{ad}

\title{Differential substitutions for non-Abelian\\ equations of KdV type}
\date{7 March 2021}
\author{V.E. Adler\thanks{L.D.~Landau Institute for Theoretical Physics, Chernogolovka, Russia. E-mail: adler@itp.ac.ru}}

\begin{document}
\maketitle

\begin{abstract}
We construct non-Abelian analogs for some KdV type equations, including the (rational form of) exponential Calogero--Degasperis equation and generalizations of the Schwarzian KdV equation. Equations and differential substitutions under study contain arbitrary non-Abelian parameters.
\medskip

\smallskip

\end{abstract}

%-------------------------------------------------------------------------------
\section{Introduction}

Equations with non-commutative (for example, matrix) variables are an important object of study in the theory of integrable systems. One of the first examples was the non-Abelian Korteweg--de Vries equation (KdV)
\[
 u_t=u_{xxx}-3uu_x-3u_xu
\]
and its modification (mKdV)
\begin{equation}\label{ft}
 f_t=f_{xxx}-3f^2f_x-3f_xf^2. 
\end{equation}
The inverse scattering method, families of exact solutions, Hamiltonian structures and Darboux--B\"acklund transformations for these equations were studied in \cite{Wadati_Kamijo_1974, Calogero_Degasperis_1977, Alonso_Olmedilla_1982, Marchenko_1986, Goncharenko_Veselov_1998, Suzko_2005} and other works. In \cite{Khalilov_Khruslov_1990}, another version of mKdV was introduced 
\begin{equation}\label{vt}
 v_t=v_{xxx}+3[v,v_{xx}]-6vv_xv, 
\end{equation}
which is related with (\ref{ft}) by an implicit change \cite{Liu_Athorne_1991}. Currently, non-Abelian analogs have been found for a large number of diverse integrable models, including evolution and hyperbolic systems, 3D equations, (semi)-discrete and ordinary differential equations. A deep, although far from complete, presentation of the subject is contained in the book \cite{Kupershmidt_2000}. A fairly wide family of nonlinear Schr\"odinger and Boussinesq type systems which are polynomial and linear with respect to the derivatives was studied in \cite{Adler_Sokolov_2021a}, and in most cases there were two non-Abelian counterparts corresponding to one scalar system, like in the case of (\ref{ft}) and (\ref{vt}). 

At the same time, there are many blank spots in the non-commutative theory, for example, no classification is known for integrable equations of the KdV type
\begin{equation}\label{eq}
 u_t=u_{xxx}+F(x,u,u_x,u_{xx}).
\end{equation}
In the scalar setting, such a classification was obtained long ago in the framework of the symmetry approach \cite{Svinolupov_Sokolov_1982, Mikhailov_Shabat_Sokolov_1991}, but in the non-Abelian case we even do not know exactly which equations from the scalar list admit a   generalization. In addition to the above equations, one can find in the literature the non-Abelian analogs of the potential mKdV equation (pmKdV) and the Schwarzian KdV equation (SKdV), see e.g. \cite[ch.~3.9]{Kupershmidt_2000} and \cite{Svinolupov_Sokolov_1996}, and that is all, up to the author's knowledge. Moreover, only homogeneous equations without parameters were studied. Meanwhile, it was recently observed \cite{Adler_Sokolov_2021b} that equation (\ref{vt}) can be generalized by adding of lower-order terms with an arbitrary non-Abelian constant, which is important when constructing a self-similar reduction and leads to a new version of the non-Abelian Painlev\'e-II equation. 

The purpose of this work is to expand the list of examples of the type (\ref{eq}) by use of differential substitutions. A similar problem was posed in \cite{Carillo_Schiavo_Porten_Schiebold_2018}, where, however, no new equations were obtained.

Recall that from the point of view of substitutions, all scalar integrable equations (\ref{eq}) are divided into three subclasses \cite{Svinolupov_Sokolov_Yamilov_1983}. One class includes the linearizable equations, the second one contains equations related to KdV, and the third one consists of one isolated Krichever--Novikov equation \cite{Krichever_Novikov_1980}
\begin{equation}\label{KN}
 z_t=z_{xxx}-\frac{3(z^2_{xx}-R(z))}{2z_x}+\alpha z_x,\quad R^{(5)}(z)=0,
\end{equation}
where $R$ is a polynomial of degree 3 or 4 with simple roots (the case of multiple roots reduces to KdV). For equations related to KdV, the basic sequence of Miura type transformations is as follows:
\begin{align}
\label{sut}
  & u_t= u_{xxx}-6uu_x,\\
\nonumber
  & \qquad \bigg\uparrow\quad u=f_x+f^2+\beta\\
\label{sft}
  & f_t= f_{xxx}-6(f^2+\beta)f_x,\\
\nonumber
  & \qquad \bigg\uparrow\quad 2f=(p_x+p^2+\alpha)/p\\
\label{spt}
  & p_t= p_{xxx}-3\frac{p_xp_{xx}}{p}+\frac{3p^3_x}{2p^2}-\frac{3}{2}\Bigl(p+\frac{\alpha}{p}\Bigr)^2p_x-6\beta p_x,\\
\nonumber
  & \qquad \bigg\uparrow\quad 2p= \bigl(\sqrt{w^2_x+4R(w)}-w_x\bigr)/(w-\gamma),\\
\label{swt}
  & w_t= w_{xxx}-\frac{3w_x(w_{xx}+2R'(w))^2}{2(w^2_x+4R(w))}+6(2w-\beta)w_x,
\end{align}
where $R(w)=(w^2-\gamma^2)(w+\gamma+\alpha)$. This ``tower of modifications'' and several simpler substitutions like the introduction of a potential cover almost all equations related to KdV (the only exception is equation (\ref{KN}) with one double root; it is related to KdV by a third-order substitution which can not be decomposed into simpler ones). Note that each modification adds an arbitrary parameter, which is important for obtaining equations of general form. In the non-Abelian setting, we can expect that some parameters may be not scalar.

One of results of this paper is the non-Abelian analog of equation (\ref{spt}) which is the rational form of the Calogero--Degasperis equation \cite{Calogero_Degasperis_1981, Fokas_1980}. It is likely that equation (\ref{swt}) (also introduced in \cite{Calogero_Degasperis_1981}) and equation (\ref{KN}) do not have non-Abelian counterparts for generic polynomials $R$. However, there exist analogs for two degenerate cases of (\ref{KN}), corresponding to $R=z^2$ and $R=1$. 

The list of non-Abelian equations and substitutions presented in the next section does not claim to be complete; it may well be that analogs of some other scalar equations can be added to this scheme. All substitutions can be verified by straightforward calculations. A more demonstrative proof based on the derivation of substitutions from auxiliary linear problems is given in section \ref{s:lambda}. 

It should be noted that the above tower of modifications for scalar equations can also be constructed based on the B\"acklund transformations, as shown by Yamilov \cite{Yamilov_1990, Yamilov_1993}, see also \cite{Borisov_Zykov_1998, Startsev_1998}. Unfortunately, this method does not work in the non-Abelian setting. However, the differential substitutions still can be used to generate the B\"acklund transformations; this is discussed briefly in section \ref{s:BT}.

%-------------------------------------------------------------------------------
\section{Graph of substitutions}\label{s:eqs}

We will construct differential substitutions between the following equations:
\begin{align}
\tag*{KdV}
 & u_t=u_{xxx}-3uu_x-3u_xu,\\[3pt]
\tag*{pKdV}
 & w_t=w_{xxx}-3w^2_x,\\[3pt]
\tag*{mKdV$_1(\alpha)$} 
 & f_t=f_{xxx}-3f^2f_x-3f_xf^2-6\alpha f_x,\\[3pt] 
\tag*{mKdV$_2(a)$} 
 & v_t=v_{xxx}+3[v,v_{xx}]-6vv_xv-3(v_x+v^2)a-3a(v_x-v^2),\\[3pt] 
\tag*{pmKdV$(a)$} 
 & y_t=y_{xxx}-3y_{xx}y^{-1}y_x-3y_xa-3yay^{-1}y_x,\\[3pt]
\tag*{CD$(a,\beta)$} 
 & p_t=(D-\ad p)\Bigl(p_{xx}-\frac{3}{2}(p_x+p^2-a)p^{-1}(p_x-p^2+a) +[p,p_x]-2p^3-6\beta p\Bigr),\\[3pt]
\tag*{CD$_0(c)$}  
 & q_t=q_{xxx}-\frac{3}{2}D\bigl((q_x-c)q^{-1}(q_x+c)\bigr),\\[3pt] 
\tag*{SKdV$(a,b,c)$} 
 & z_t=z_{xxx}-\frac{3}{2}(z_{xx}-az+zb-c)z^{-1}_x(z_{xx}+az-zb+c) -3az_x-3z_xb.
\end{align}
Here $D$ denotes the derivative with respect to $x$, field variables $u,\dots,z$ and constants $a,b,c$ belong to a free associative algebra ${\mathcal A}$ over $\mathbb C$ with unit 1, and $u^{-1}$ denote the inverse element for $u$. For the sake of clarity, one can assume that $\mathcal A$ is the algebra of matrices of some arbitrary fixed size. Equations mKdV$_1(\alpha)$ and CD$(a,\beta)$ also contain scalar constants $\alpha,\beta\in\mathbb C$. These parameters can be set to $0$ by the Galilean transform, but we do not do this, since they are involved in the substitutions and related B\"acklund transformations. In what follows, we identify $\alpha$ with $\alpha 1\in\mathcal A$, which gives meaning for expressions like $u+\alpha$. 

For any equation, the order of factors in monomials can be reversed by passing to the new operation of multiplication $a\circ b=ba$ on ${\mathcal A}$. The resulting equation will be called transposed by analogy with the matrix case. It is natural to consider it equivalent to the original equation. The above equations have the property of invariance with respect to the transposition, possibly up to some additional involutions like the sign change (in pmKdV, we apply $y\to y^{-1}$) or changing the parameters.

\begin{remark}
In the scalar case, equation CD$(\alpha,\beta)$ coincides with (\ref{spt}), the rational form of Calogero--Degasperis equation. Passing to the exponents brings it and CD$_0(\gamma)$ to the original form from \cite{Calogero_Degasperis_1981, Fokas_1980}:
\begin{gather*}
 P_t=P_{xxx}-\frac{1}{2}P^3_x-\frac{3}{2}(e^P+\alpha e^{-P})^2P_x-6\beta P_x,\quad p=e^P,\\
 Q_t=Q_{xxx}-\frac{1}{2}Q^3_x-\frac{3\gamma^2}{2}e^{-2Q}Q_x,\quad q=e^Q.
\end{gather*}
The same change brings the scalar equation pmKdV$(\alpha)$ to the usual polynomial form
\[
 Y_t=Y_{xxx}-2Y^3_x-6\alpha Y_x,\quad y=e^Y.
\] 
\end{remark}

\begin{remark}
Comparing the above equations for $Q$ and $Y$, we see that CD$_0(0)$ is actually another non-Abelian analog of pmKdV. This equation was studied in \cite{Svinolupov_Sokolov_1996}. One more interesting analog of pmKdV on Jordan algebras was obtained in \cite[(2.4)]{Svinolupov_1993}. In contrast to CD$_0(0)$, it involves the operator $M^{-1}_q$ instead of $q^{-1}$, where $M_q$ is the multiplication operator in the Jordan algebra. Since any associative algebra ${\mathcal A}$ turns into a Jordan algebra with respect to the operation $a\circ b=\tfrac{1}{2}(ab+ba)$, such equation can be defined on ${\mathcal A}$ as well, if we allow expressions involving $(L_q+R_q)^{-1}$, where $L_q$ and $R_q$ are operators of the left and right multiplication in ${\mathcal A}$.
\end{remark}

\begin{remark}
Scalar equation SKdV$(a,b,c)$ is actually the degenerate case of the Krichever--Novikov equation (\ref{KN}) with $R=((a-b)z+c)^2$. The case $R=0$ is the Schwarzian KdV equation, and we keep this name for the whole family of equations.
\end{remark}

The following proposition is the main result of the paper.

\begin{proposition}\label{prop:sub}
The above non-Abelian equations are related by substitutions\upshape
\begin{alignat}{2}
\label{wu}
 \text{pKdV}&\to\text{KdV}: &\qquad& u=w_x,\\
\label{yf}
 \text{pmKdV}(\alpha)&\to\text{mKdV}_1(\alpha): &\qquad& f=y_xy^{-1},\\
\label{fu}
 \text{mKdV}_1(\alpha)&\to\text{KdV}: && u=f_x+f^2+\alpha,\\
\label{yu}
 \text{pmKdV}(a)&\to\text{KdV}: && u=y_{xx}y^{-1}+yay^{-1},\\
\label{yv} 
 \text{pmKdV}(a)&\to\text{mKdV}_2(a): && v=y^{-1}y_x,\\
\label{zp}
 \text{SKdV}(a,\beta,0)&\to\text{CD}(a-\beta,\beta): &&  p=z_xz^{-1},\\
\label{pv}
 \text{CD}(a-\beta,\beta)&\to\text{mKdV}_2(a): && v=-\tfrac{1}{2}(p_x+p^2-a+\beta)p^{-1},\\
\label{zq}
 \text{SKdV}(0,0,c)&\to\text{CD}_0(c): && q=z_x,\\
\label{qv}
 \text{CD}_0(c)&\to\text{mKdV}_2(0): && v=-\tfrac{1}{2}(q_x-c)q^{-1},\\
\label{zv}
 \text{SKdV}(a,b,c)&\to\text{mKdV}_2(a): && v=-\tfrac{1}{2}(z_{xx}-az+zb-c)z^{-1}_x.
\end{alignat}
\end{proposition}

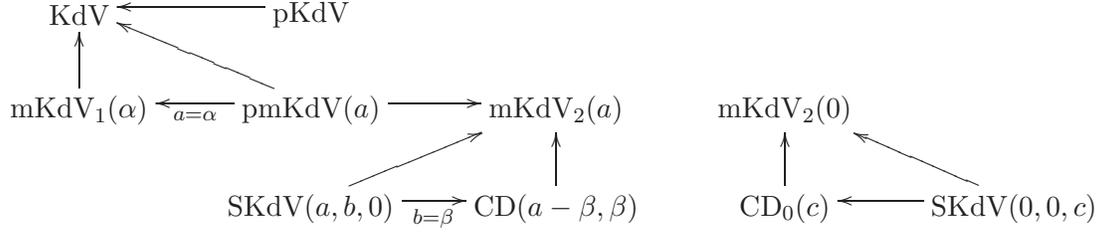
\begin{figure}[t]
\centerline{\xymatrix@R=20pt@C=24pt{
 \txt{KdV} & \txt{pKdV} \ar[l] \\
 \txt{mKdV$_1(\alpha)$} \ar[u] & \txt{pmKdV$(a)$} \ar[l]^{a=\alpha} \ar[r] \ar[lu] & \txt{mKdV$_2(a)$} & \txt{mKdV$_2(0)$} \\ 
 &  \txt{SKdV$(a,b,0)$} \ar[r]_{b=\beta} \ar[ur] & \txt{CD$(a-\beta,\beta)$} \ar[u] & 
 \txt{CD$_0(c)$} \ar[u] & \txt{SKdV$(0,0,c)$} \ar[ul] \ar[l]}} 
\caption{Substitutions chart.}\label{fig:subst}
\end{figure}

Schematically, these substitutions are shown in Fig.~\ref{fig:subst}. Note that the substitution (\ref{yf}) from pmKdV$(a)$ is valid only for the scalar value of parameter $a=\alpha$, so the substitution (\ref{yu}) is more general than the composition of (\ref{yf}) and (\ref{fu}). Similarly, the substitutions (\ref{zp}) and (\ref{zq}) from SKdV$(a,b,c)$ do not work for arbitrary $(a,b,c)$, so the substitution (\ref{zv}) is more general than the compositions of (\ref{zp}) and (\ref{pv}) or of (\ref{zq}) and (\ref{qv}).

It is worth noting that if we consider the complete graph of substitutions for scalar equations (the sequence (\ref{swt}) $\to\dots\to$ (\ref{sut}) from Introduction is a part of this graph) then the KdV equation (\ref{sut}) turns out to be the only vertex without outgoing arrows, leaving apart the isolated Krichever--Novikov equation (\ref{KN}). In the scalar setting, we can say that all paths lead to KdV. This is not the case for non-Abelian equations, since the mKdV$_2$ equation also turns out to be an end-point vertex.
 
%-------------------------------------------------------------------------------
\section{Derivation of substitutions from linear problems}\label{s:lambda}

The auxiliary linear equations for the non-Abelian KdV equation look the same as in the scalar case:
\[
 \psi_{xx}=(u-\lambda)\psi,\quad \psi_t=u_x\psi-(2u+4\lambda)\psi_x.
\]
However, this can be generalized by replacing the scalar spectral parameter $\lambda$ with a non-Abelian one:
\begin{equation}\label{psieqs}
 \psi_{xx}=u\psi-\psi\Lambda,\quad \psi_t=u_x\psi-2u\psi_x-4\psi_x\Lambda,
\end{equation}
where $\psi,\Lambda\in{\mathcal A}$. Indeed, here $\Lambda$ acts on $\psi$ as the operator of right multiplication $R_\Lambda$, and the coefficients $u$ and $u_x$ act as operators of left multiplication $L_u$ and $L_{u_x}$. But, any operators of left and right multiplication commute: $R_aL_b=L_bR_a$, $a,b\in{\mathcal A}$, therefore $R_\Lambda$ behaves exactly like the scalar coefficient $\lambda$ when calculating the compatibility condition. It is clear that such a generalization is possible for any zero curvature representation $A_t=B_x+[B,A]$, where entries of the matrices $A$ and $B$ are elements of $\mathcal A$ depending on $\lambda$ and field variables. 

We use (\ref{psieqs}) as the starting point for deriving the substitutions from the Proposition~\ref{prop:sub}.
\smallskip

{\it PmKdV equation.} As in the scalar case, the Miura transform is constructed by a particular solution $\psi=y$ corresponding to a fixed value $\Lambda=a\in{\mathcal A}$:
\begin{equation}\label{yeqs}
 y_{xx}=uy-ya,\quad y_t=u_xy-2uy_x-4y_xa.
\end{equation}
From the first equation we find $u=y_{xx}y^{-1}+yay^{-1}$ and the elimination of $u$ from the second equation yields pmKdV$(a)$. Thus, pmKdV is the equation for the wave function of the Schr\"odinger operator, and the substitution (\ref{yu}) is a rewrite of the original linear equation.
\smallskip

{\it Equations mKdV$_1$ and mKdV$_2$.} The next step in the scalar setting is to pass to the logarithmic derivative $f=y_x/y$. With non-commutative variables, this can be done in at least two ways, leading to different answers.

The substitution $f=y_xy^{-1}$ works only for $a=\alpha\in\mathbb C$. In this case, the first equation (\ref{yeqs}) gives the Miura transformation $u=f_x+f^2+\alpha$ which is the substitution (\ref{fu}), and the second equation gives 
\begin{equation}\label{yF}
 y_t=(u_x-2uf-4\alpha f)y=Fy,\quad F=f_{xx}+[f,f_x]-2f^3-6\alpha f.
\end{equation}
From equations $y_x=fy$, $y_t=Fy$ it follows $f_t=(D-\ad f)(F)$, which coincides with mKdV$_1(\alpha)$.

If we apply another change $v=y^{-1}y_x$, the first equation (\ref{yeqs}) takes the form $y^{-1}uy=v_x+v^2+a$. By replacing $u$ in the second equation, we obtain
\[
 y_t=y(y^{-1}u_xy-2y^{-1}uyv-4va)=yV,\quad V=v_{xx}+2[v,v_x]-2v^3-3av-3va.
\]
From equations $y_x=yv$ and $y_t=yV$ it follows $v_t=(D+\ad v)(V)$ which coincides with mKdV$_2(a)$.
\smallskip

{\it Equation SKdV.} The next step is to transform equations (\ref{psieqs}) into auxiliary linear problems for mKdV$_2$, in such a way that the potential $u$ is replaced by $v$. To do this it is enough to make the gauge transformation $\psi=y\varphi$ \cite{Liu_Athorne_1991}. In other words, we define $\varphi$ as the ratio of solutions for (\ref{psieqs}) and (\ref{yeqs}). A direct calculation leads to equations
\begin{equation}\label{phieqs}
 \varphi_{xx}=a\varphi-\varphi\Lambda-2v\varphi_x,\quad
 \varphi_t= 4v(a\varphi-\varphi\Lambda)-2(v_x+v^2+a)\varphi_x-4\varphi_x\Lambda,
\end{equation}
where $\varphi,a,\Lambda\in{\mathcal A}$. By construction, the compatibility condition for this pair of equations is equivalent to mKdV$_2(a)$. Now we can do what we did before with the pair (\ref{psieqs}): to write an equation for a particular solution $\varphi=z$ corresponding to the value $\Lambda=b$ and for its logarithmic derivative. So, let $z$ satisfy the equations
\begin{equation}\label{zxx}
 z_{xx}=az-zb-2vz_x,\quad z_t=4v(az-zb)-2(v_x+v^2+a)z_x-4z_xb.
\end{equation}
We find $v$ from the first equation and substitute it into the second one. After simple algebra, this leads to the SKdV$(a,b,0)$ equation
\[
 z_t=z_{xxx}-\frac{3}{2}(z_{xx}-az+zb)z^{-1}_x(z_{xx}+az-zb)-3az_x-3z_xb
\]
and to the substitution $-2v=(z_{xx}-az+zb)z^{-1}_x$ which relates this equation with mKdV$_2(a)$. 

In this equation, it is easy to see that the transformation $z\to z+d$ leads only to the replacement of the term $az-zb$ with $az-zb+c$, where $c=ad-db$. For arbitrary $a,b,d\in\mathcal A$, the element $c$ is arbitrary as well, therefore the equation can be extended to the case of three parameters SKdV$(a,b,c)$. This brings to the substitution (\ref{zv}).
\smallskip

{\it Equations CD$(a,\beta)$ and CD$_0(c)$.} Since equation SKdV$(0,0,c)$ contains only derivatives of $z$, it admits the substitution $z_x=p$ which leads to CD$_0(c)$. For equation SKdV$(a,b,0)$, we can use the logarithmic derivative instead, assuming that one of parameters $a$ or $b$ is scalar. In contrast to the pmKdV equation for $y$, this equation is invariant with respect to the transposition and the interchange of parameters, so that both versions $z_xz^{-1}$ and $z^{-1}z_x$ are on equal footing. For definiteness, let
\[
 p=z_xz^{-1},\quad b=\beta\in\mathbb C,
\]
then equations (\ref{zxx}) take the form
\begin{equation}\label{zP}
 p_x+p^2=a-\beta-2vp,\quad z_tz^{-1}=4v(a-\beta)-2(v_x+v^2+a+2\beta)p.
\end{equation}
The first equation is equivalent to the substitution (\ref{pv}), and the second one takes the form $z_t=Pz$ after eliminating $v$, with
\[
 P=p_{xx}-\frac{3}{2}(p_x+p^2-a+\beta)p^{-1}(p_x-p^2+a-\beta) +[p,p_x]-2p^3-6\beta p.
\]
From here the equation $p_t=(D-\ad p)(P)$ follows, which is CD$(a-\beta,\beta)$.

%-------------------------------------------------------------------------------
\section{B\"acklund transformations}\label{s:BT}

Recall that the scalar tower of modifications (\ref{sut})--(\ref{swt}) can be derived in an alternative way based on B\"acklund transformations \cite{Yamilov_1990,Yamilov_1993}. For the mKdV equation, the $x$-part of B\"acklund transformations is represented by the dressing chain \cite{Veselov_Shabat_1993}
\begin{equation}\label{ff}
 f_{n,x}+f_{n+1,x}=f^2_n-f^2_{n+1}+\alpha_n-\alpha_{n+1}
\end{equation}
generated by compositions of the Miura map and the change $f\to -f$ which leaves the mKdV equation invariant. Let us introduce a new variable $p_n$ and rewrite (\ref{ff}) as the system
\[
 f_n+f_{n+1}=p_n,\quad f_n-f_{n+1}=\frac{p_{n,x}-\alpha_n+\alpha_{n+1}}{p_n},
\]
then both $f_n$ and $f_{n+1}$ are easily expressed through $p_n$ and $p_{n,x}$. This is the substitution (\ref{spt})$\to$(\ref{sft}), up to the values of parameters, and moreover, we obtain the chain for $p_n$ 
\begin{equation}\label{pp}
 (p_np_{n+1})_x=p_np_{n+1}(p_n-p_{n+1}) +(\alpha_n-\alpha_{n+1})p_{n+1}+(\alpha_{n+1}-\alpha_{n+2})p_n.
\end{equation}
The CD  equation for $p_n$ is obtained by applying the found substitutions to the result of differentiating the relation $p_n=f_n+f_{n+1}$ in virtue of the mKdV equations for $f_n$ and $f_{n+1}$. Moreover, the invariance with respect to the shift $n\to n+1$ guarantees that $p_{n+1}$ also satisfies the CD equation, so that the chain (\ref{pp}) defines the $x$-part of B\"acklund transformations for this equation. 

This trick can be repeated once more by introducing the variable $w_n=p_np_{n+1}$ and this leads to the substitution (\ref{swt})$\to$(\ref{spt}). Further modification is not possible, since the chain equation for $w_n$ does not have some functional structure required for this, see \cite{Yamilov_1993}.

What about this procedure in the non-Abelian setting? One can easily see that the chain (\ref{ff}) still works, since the form of the Miura map (\ref{fu}) is the same as for the scalar counterparts. Therefore, (\ref{ff}) defines the B\"acklund transformations for the non-Abelian mKdV$_1(\alpha)$ equation. However, introducing the variable $p_n=f_n+f_{n+1}$ now gives nothing for the simple reason that the difference of non-Abelian squares $f^2_n-f^2_{n+1}$ is not factorizable. Unfortunately, this nice scheme breaks down already at the first step.

Nevertheless, the chain of B\"acklund transformations for the non-Abelian CD equation can still be constructed, but for this we have to use the substitution (\ref{pv}) from CD to mKdV$_2$ and the change $p\to-p^t$ which leaves CD equation invariant. Hence it follows that there are two different substitutions between the CD$(a-\beta,\beta)$ and the mKdV$_2(a)$ equations:
\[
 -2v=p^{-1}(-p_x+p^2-a+\beta)\qquad \text{and} \qquad -2v=(p_x+p^2-a+\beta)p^{-1}.
\]
In these substitutions, $\beta$ can be changed, but $a$ is fixed, since this parameter is contained in the target mKdV$_2(a)$ equation. This gives rise to a sequence of substitutions
\[
 -2v_n=p^{-1}_n(-p_{n,x}+p^2_n-a+\beta_n)=(p_{n+1,x}+p^2_{n+1}-a+\beta_{n+1})p^{-1}_{n+1},
\]
from which we obtain the chain
\begin{equation}\label{pp'}
 (p_np_{n+1})_x = p_n(p_n-p_{n+1})p_{n+1} -(a-\beta_n)p_{n+1} +p_n(a-\beta_{n+1}).
\end{equation}
By construction, it defines the $x$-part of B\"acklund transformations between equations CD$(a-\beta_n,\beta_n)$. Of course, this remains true also for scalar variables, but note that the scalar version of (\ref{pp'}) {\em differs from} (\ref{pp}): in one equation the signs of two linear terms coincide, and in the other one they are opposite. This distinction is due to different methods which we used to construct the chains, and it turns out that one method admits a non-Abelian generalization, while the other does not.

In conclusion, we remark that the Miura maps (\ref{fu}) and (\ref{pv}) admit non-local generalizations. Relations like (\ref{yF}), which were obtained under assupmtion $a=\alpha\in\mathbb C$, can be obtained also for $a\in\mathcal A$, at the expense of introducing an additional variable. Let $f=y_xy^{-1}$, $g=yay^{-1}$ then relations (\ref{yeqs}) imply
\[
 u=f_x+f^2+g,\quad y_t=Fy,\quad F=f_{xx}+[f,f_x]-2f^3-3fg-3gf,
\] 
and we arrive to the following system for $f$ and $g$:
\[
 f_t=(D-\ad f)(F)=f_{xxx}-3(f^2+g)f_x-3f_x(f^2+g),\quad g_t=[F,g],\quad g_x=[f,g].
\]
The variable $g$ can be viewed as a nonlocality defined by the latter equation which plays the role of a constraint. The invariance of all relations with respect to the change $(u,f,g)\to(u^t,-f^t,g^t)$ implies that there is also another Miura map $u=-f_x+f^2+g$. The composition of these two substitutions gives rise to the dressing chain
\[
 f_{n,x}+f_{n+1,x}=f^2_n-f^2_{n+1}+g_n-g_{n-1},\quad g_{n,x}=[f_n,g_n],
\]
which defines the general Darboux transformations for the non-Abelian Schr\"odinger operator \cite{Suzko_2005} and the $x$-part of B\"acklund transformations for the above system for $f$ and $g$. 

In a similar way, relations (\ref{zP}) obtained under assumption $b=\beta\in\mathbb C$ can be generalized for the case $b\in\mathcal A$, by introducing the additional variable $h=zbz^{-1}$. This leads to the system
\begin{gather*}
 p_t=(D-\ad p)(P),\quad h_t=[P,h],\quad h_x=[p,h],\\
 P=p_{xx}-\frac{3}{2}(p_x+p^2-a+h)p^{-1}(p_x-p^2+a-h)+[p,p_x]-2p^3-3ph-3hp
\end{gather*}
and to the chain
\[
 (p_np_{n+1})_x = p_n(p_n-p_{n+1})p_{n+1} -(a-h_n)p_{n+1} +p_n(a-h_{n+1}),\quad h_{n,x}=[p_n,h_n],
\]
which defines the $x$-part of B\"acklund transformations for this system and the Darboux transformations for the spectral problem (\ref{phieqs}).

%-------------------------------------------------------------------------------
\subsubsection*{Acknowledgments}

Work supported by the Russian Foundation for Basic Research grant \#\,20-52-05015 Arm\_a.

%-------------------------------------------------------------------------------

\end{document}